\documentclass[sigconf]{acmart}

\AtBeginDocument{%
  }

\usepackage{xcolor}
\usepackage{lipsum}

\begin{document}

\title{RBoard: A Unified Platform for Reproducible and Reusable Recommender System Benchmarks}

\author{Xinyang Shao}
\affiliation{
  \institution{Huawei Ireland Research Centre}
  \city{Dublin}
  \country{Ireland}}
\email{xinyang.shao@huawei-partners.com}

\author{Edoardo D'Amico}
\affiliation{
  \institution{Huawei Ireland Research Centre}
  \city{Dublin}
  \country{Ireland}}
\email{edoardo.damico@huawei-partners.com}

\author{Gabor Fodor}
\affiliation{
  \institution{Huawei Ireland Research Centre}
  \city{Dublin}
  \country{Ireland}}
\email{gabor.fodor@huawei-partners.com}

\author{Tri Kurniawan Wijaya}
\affiliation{
  \institution{Huawei Ireland Research Centre}
  \city{Dublin}
  \country{Ireland}}
\email{tri.kurniawan.wijaya@huawei.com}


\begin{abstract}
Recommender systems research lacks standardized benchmarks for reproducibility and algorithm comparisons. We introduce RBoard, a novel framework addressing these challenges by providing a comprehensive platform for benchmarking diverse recommendation tasks, including CTR prediction, Top-N recommendation, and others. RBoard's primary objective is to enable fully reproducible and reusable experiments across these scenarios. The framework evaluates algorithms across multiple datasets within each task, aggregating results for a holistic performance assessment. It implements standardized evaluation protocols, ensuring consistency and comparability. To facilitate reproducibility, all user-provided code can be easily downloaded and executed, allowing researchers to reliably replicate studies and build upon previous work. By offering a unified platform for rigorous, reproducible evaluation across various recommendation scenarios, RBoard aims to accelerate progress in the field and establish a new standard for recommender systems benchmarking in both academia and industry. The platform is available at \url{https://rboard.org} and the demo video can be found at \url{https://bit.ly/rboard-demo}.

\end{abstract}

\ccsdesc[500]{Information systems~Recommender systems}
\ccsdesc[300]{Information systems~Evaluation of retrieval results}

\keywords{Recommender Systems, Evaluation Framework, Benchmarking, Reproducibility.}

\received{20 February 2007}
\received[revised]{12 March 2009}
\received[accepted]{5 June 2009}

\maketitle

\section{Introduction}
Recommender systems play a crucial role in shaping our digital experiences, from product discovery to content consumption \cite{ricci2010introduction}. As this field rapidly evolves, the need for standardized evaluation methods and replicable experiments has become critical. Despite algorithmic advancements, the research community struggles to effectively compare approaches and validate results across diverse tasks and datasets \cite{said2014comparative, beel2016towards, dacrema2019are}.

The recommender systems domain encompasses a wide range of tasks, including click-through rate (CTR) prediction, top-N recommendation, and sequential and session-based recommendations \cite{yang2022click, cremonesi2010performance, wang2021survey}. Each of these tasks presents unique challenges and requires specific evaluation metrics. However, the lack of a unified benchmarking framework has led to inconsistent evaluation protocols \cite{zhao2022revisiting}, limited reproducibility \cite{dacrema2019are}, and potential dataset biases. These challenges impede progress on two fronts: researchers struggle to build upon each other's work, while companies face difficulties in identifying and integrating the most effective algorithms for their unique business needs.

To address these issues, we introduce RBoard, a novel framework that offers a systematic approach to evaluating and comparing recommender systems. RBoard distinguishes itself by providing a unified platform for benchmarking diverse recommender tasks, with a primary focus on enabling fully reproducible and reusable experiments. The framework implements consistent evaluation protocols across multiple datasets for each task, ensuring fair comparisons and offering a comprehensive assessment of algorithm performance. By facilitating downloading and execution of user-provided code, RBoard streamlines experiment replication and promotes transparency in the research process. This approach aims to establish a new standard for benchmarking recommender systems algorithms.

\section{Key features and Innovations}
\begin{figure*}
    \centering
    \includegraphics[width=0.9\linewidth]{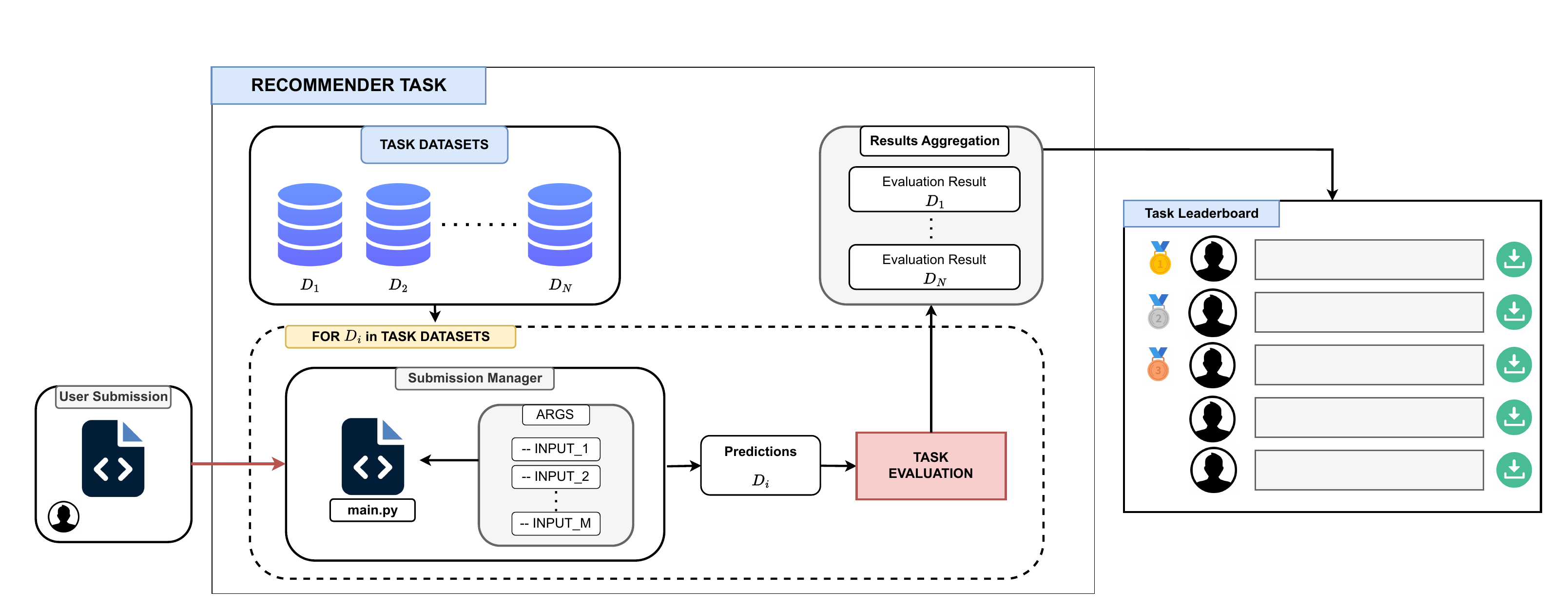}
    \caption{Architecture overview of the RBoard Framework.}
    \label{fig:rboard-schema}
\end{figure*}
Recommender system research is shaped by two main tools: full-pipeline frameworks and benchmarking platforms. Full-pipeline frameworks like Elliot \cite{anelli2021elliot}, RecBole \cite{zhao2021recbole}, and SSLRec \cite{ren2024sslrec} offer comprehensive solutions for dataset preprocessing, algorithm implementation, and evaluation. While these frameworks have advanced standardized workflows, challenges in ensuring reproducibility and reusability persist due to variations in implementation details and the need for modifications in complex experiments.
Existing benchmarking platforms \cite{DBLP:conf/sigir/ZhuDSMLCXZ22} often fall short in providing truly reproducible benchmarks by overlooking critical aspects of the recommendation process, particularly hyperparameter tuning. This oversight limits the practical relevance of benchmarks and fails to capture important real-world considerations, such as the time required for hyperparameter optimization.
RBoard addresses these challenges by introducing a flexible, task-agnostic benchmarking environment that accommodates a wide range of recommender system scenarios and frameworks. This versatility allows RBoard to complement existing tools while focusing on two core objectives:
\begin{itemize}
\item \textbf{Reproducibility:} Ensuring that all experiments can be easily replicated, regardless of the underlying implementation approach.
\item \textbf{Reusability:} Facilitating the reuse of research code and methodologies across different studies and contexts.
\end{itemize}
RBoard's design accommodates various implementation approaches, from custom code to full-pipeline frameworks, while providing a standardized evaluation environment. This flexibility allows researchers to use their preferred methods while still benefiting from consistent benchmarking. By enabling direct comparison of experiments and enhancing the reusability of research code, RBoard aims to improve the reproducibility of results across diverse recommender system studies.

\section{System Design}
RBoard's architecture is fundamentally designed to ensure reproducibility and reusability in recommender system research. This is achieved through the following design choices:

\noindent
\textbf{Data Handling and Preprocessing.}
Central to RBoard's design is its approach to data handling and preprocessing. The platform manages dataset preprocessing and splitting, ensuring consistency across experiments and eliminating variations in data preparation that could affect reproducibility. Users receive preprocessed input data for model training and prediction, while test sets remain hidden to maintain evaluation integrity. To foster transparency, RBoard makes preprocessing code available for review, but withholds specific randomization element to prevent overfitting and ensure result generalizability.

\noindent
\textbf{Task Evaluation.}
The evaluation process provides task-specific metrics with openly available evaluation code, allowing researchers to understand precisely how their algorithms are assessed. A key innovation of RBoard is its aggregation of results across multiple datasets within each task, offering a more comprehensive and generalizable view of algorithm performance. This multi-dataset evaluation approach helps mitigate dataset-specific biases and provides a more robust assessment of recommender systems across varied contexts.

\noindent
\textbf{User Code Integration.}
User code integration follows a standardized approach. Researchers upload their code with a main.py file as the entry point, which the platform executes with necessary inputs as command-line arguments. This uniform method ensures consistency across submissions and simplifies experiment reproduction.

\noindent
\textbf{Hyper-parameters tuning.}
RBoard identifies hyperparameter tuning as an essential component of reproducible experiments. Users must include tuning processes in their main.py file, ensuring the entire experimental pipeline can be replicated. This approach not only captures the final performance metrics but also assesses the time and computational resources required for optimization.
 

\noindent
\textbf{Code Availability and Reusability.}
To promote reusability and collaboration, RBoard makes all submitted code available for download, allowing researchers to build upon existing work and verify results independently.

\section{Workflow And Use cases}
As illustrated in \autoref{fig:rboard-schema}, RBoard's workflow begins with users uploading their code, including required hyperparameter tuning. The submission manager executes this code for each dataset within a task, collecting and evaluating predictions on separate test data. RBoard then aggregates results to compute overall performance metrics. Both aggregated and single dataset results are displayed on a public leaderboard, offering a comprehensive view of algorithm effectiveness across diverse data contexts.
 
RBoard serves diverse use cases in the recommender systems community. Researchers can benchmark their algorithms against state-of-the-art approaches, gaining insights into performance across various scenarios. Industry practitioners can identify top-performing algorithms and adapt them to their specific needs, bridging the gap between academic research and real-world applications.


\section{Conclusions and Future Works}
RBoard addresses critical challenges in recommender system research by providing a standardized platform for benchmarking diverse tasks. Its key contributions include a unified evaluation environment, standardized protocols, and open code availability, all enhancing reproducibility and reusability. Future work will focus on expanding the range of supported tasks, exploring different splitting and preprocessing protocols, and developing diverse strategies for aggregating benchmark results across datasets. 



\bibliographystyle{ACM-Reference-Format}
\bibliography{main}


\end{document}